\documentclass[a4paper, 12pt]{article}

\usepackage{graphicx}
\usepackage{graphicx}

\usepackage{amsfonts,amsmath}
\usepackage{graphicx}
\usepackage{authblk}
\usepackage{hyperref}
\usepackage{microtype}
\usepackage{lipsum}
\usepackage{afterpage}
\usepackage{capt-of}
\usepackage{pifont}
\usepackage{bm}
\usepackage{pdfpages}
\usepackage{mdframed}

\def\vol{\mbox{vol}}

\def\RR{{\mathbb R}}

\usepackage{geometry}
\begin{document}
\date{\quad}
\title{Modeling of crisis periods in stock markets}
\author[1,2]{Apostolos Chalkis}
\author[1,2]{Emmanouil Christoforou}
\author[2]{Theodore Dalamagkas}
\author[1,2]{Ioannis Z.~Emiris}

\affil[1]{Department of Informatics \& Telecommunications\\ National \& Kapodistrian University of Athens, Greece}
\affil[2]{ATHENA Research \& Innovation Center, Greece}

%
%
%

%
\maketitle              
\begin{abstract}
We exploit a recent computational framework to model and detect financial crises in stock markets, as well as shock events in cryptocurrency markets, which are characterized by a sudden or severe drop in prices. 
Our method manages to detect all past crises in the French industrial stock market starting with the crash of 1929, including financial crises after 1990 (e.g.\ dot-com bubble burst of 2000, stock market downturn of 2002), and all past crashes in the cryptocurrency market, namely in 2018, and also in 2020 due to covid-19.
We leverage copulae clustering, based on the distance between probability distributions, in order to validate the reliability of the framework; we show that clusters contain copulae from similar market states such as normal states, or crises.
Moreover, we propose a novel regression model that can detect successfully all past events using less than $10\%$ of the information that the previous framework requires.
We train our model by historical data on the industry assets, and we are able to detect all past shock events in the cryptocurrency market.
Our tools provide the essential components of our software framework that offers fast and reliable detection, or even prediction, of shock events in stock and cryptocurrency markets of hundreds of assets.
 
\end{abstract}

\section{Introduction}

Modern finance has been pioneered by Markowitz who set a framework to study choice in portfolio allocation under uncertainty \cite{M52}.
Within this framework, portfolios are characterized by their returns, and by their risk which is defined as the variance (or volatility) of the portfolios' returns.
An investor would build a portfolio to maximize its expected return for a chosen level of risk. 
In normal times, stocks are characterized by somewhat positive returns and a moderate volatility, in up-market times (typically bubbles), by high returns and low volatility, and during financial crises, by strongly negative returns and high volatility~\cite{BGP12}.
Thus, it is crucial to describe the time-varying dependency between portfolios' returns and volatility.
\begin{figure}[t]
 \begin{minipage}[h]{0.34\textwidth}
\includegraphics[width=\linewidth]{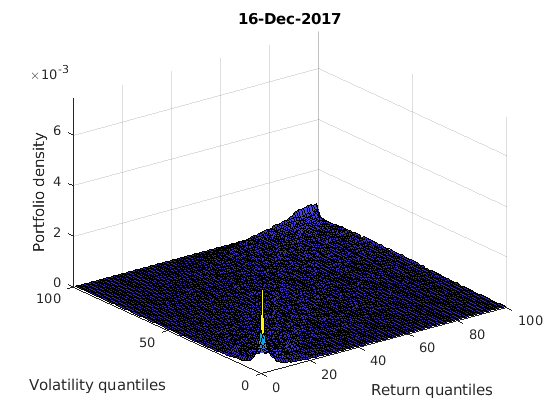}
\end{minipage}
 \begin{minipage}[h]{0.3\textwidth}
\includegraphics[width=\linewidth]{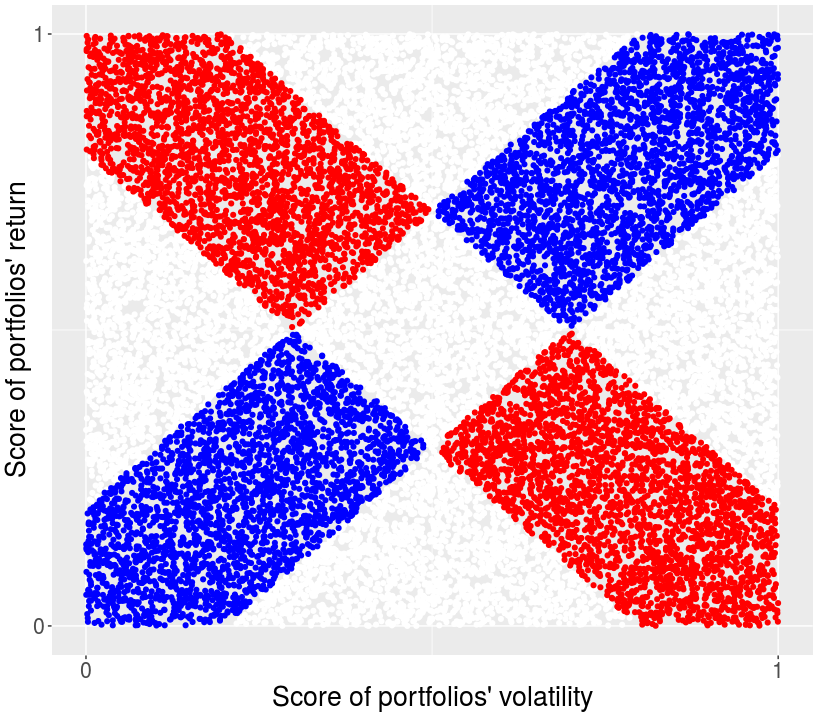}
\end{minipage}
\begin{minipage}[h]{0.34\textwidth}
\includegraphics[width=\linewidth]{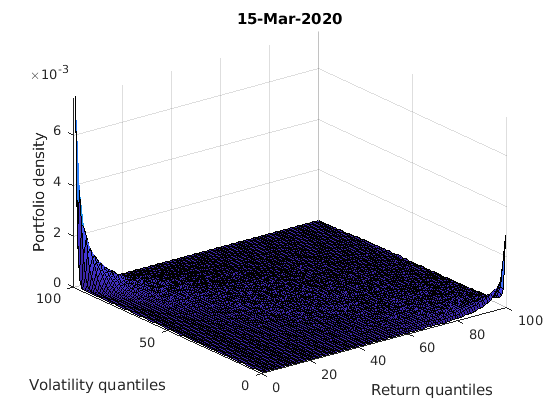}
\end{minipage}
\caption{Copulae that correspond to cryptocurrencies' states. Left, a normal period (16/12/2017) and right, a shock event due to Covid-19 (15/03/2020). The middle plot shows the mass of interest to characterize the market state.\label{fig:cop2}}
\end{figure}
To capture this dependency, in~\cite{Cales18} they rely on the copula representation of the portfolios distribution, which offers a very powerful tool.
A copula is a bivariate probability distribution for which the marginal probability distribution of each variable is uniform.
Following Markowitz' framework, the variables considered are the portfolios' return and volatility.
Fig.~\ref{fig:cop2} illustrates a copula showing a positive (left) and negative (right) dependency between return and variance. 



We illustrate the power of the framework in~\cite{Cales18} to detect shocks in two different markets: the French industrial stock market and the market of digital assets. We use the daily returns of 30 French industrial assets to detect all reported financial crises after 1990. We also detect earlier crises, such as the crash of 1929 (Fig.~\ref{fig:WarningsCrises}).  Interestingly, the indicator recognizes the period of military occupation of France (1940-45) as normal; this is related to the strict regulation during that period, which led to a paradoxical rise in nominal stock values~\cite{oosterlinck10,bris2012}. 
Also, we use the daily returns of $12$ cryptocurrencies with the longest history available to detect all shock events in the cryptocurrency market. The indicator detects successfully the 2018 cryptocurrency crash and the timeline of its most notable events, such as the crash of nearly all cryptocurrencies in the beginning of 2018 and the fall of Bitcoin's market capitalization and price in the end of 2018. Finally, it detects the shock event in early 2020 due to covid-19.

We validate the reliability of the framework by clustering based on probability distributions' distances. The computed copulae form clusters that sort the value of the indicator, resulting to clusters of similar financial states (normal, crisis). 
 We employ quadratic regression models to model the copula structure so as to capture several patterns of the mass of portfolios during different market states. We train our model using the French industry assets and we use it to detect shock events in the cryptocurrency market. Our trained model successfully detects all past events. It uses less than $10\%$ of the information on the dependency between portfolios' return and volatility that is required in~\cite{Cales18}. Lastly, the open-source implementation\footnote{\scriptsize\url{https://github.com/TolisChal/crises_detection}} of our methods provides a software framework for shock event detection and modeling in stock markets of hundreds of assets.

The rest of the paper is organized as follows. Section~\ref{sec:framework} presents the computational framework~\cite{Cales18} and uses real data to detect past crises and shock events. Section~\ref{sec:dynamic_copulas} exploits clustering and regression models to provide more sophisticated tools on crisis detection and modeling.
 
\section{Detecting shock events with copulae}\label{sec:framework}

In this section we present the computational framework in~\cite{Cales18} and then we exploit it to detect past crises and crashes in two markets with different characteristics.


Let a portfolio $x$ invest in $n$ assets. The set of portfolios in which a long-only asset manager can invest can be represented by the canonical simplex 
$$
\Delta^{n-1} := \left\{ (x_1, \dots, x_n) \in \mathbb{R}^{n}\ \left|\ \sum_{i=1}^{n} x_i=1, \mbox{ and } x_i \ge 0\} \right.\right\}\subset \RR^n .
$$
Given a vector of asset returns $R\in\RR^n$ and the variance-covariance matrix $\Sigma\in\RR^{n\times n}$ of the distribution of asset returns, we say that any portfolio $x \in \Delta^{n-1}$ has return $f_{ret}(x, R) = R^T x$ and variance (volatility) $f_{vol}(x, \Sigma) =  x^T\Sigma x$. 

To capture the relationship between return and volatility in a given time period we approximate the {\em copula} between portfolios' return and volatility. Thus, we define two sequences of $m$ bodies each, $\Delta^{n-1}\cap S_i:=\{ x\in \Delta^{n-1}\ |\ s_i\leq f_{ret}(x, R)\leq s_{i+1}\}$ and $\Delta^{n-1}\cap U_i:=\{ x\in\Delta^{n-1}\ |\ u_i\leq f_{vol}(x, \Sigma)\leq u_{i+1}\},\ i\in[m]$. Moreover, we compute $s_i,\ u_i\in\RR$ 
such that
$\Delta^{n-1}\cap\vol(S_i)$ and $\Delta^{n-1}\cap\vol(U_i)$ are all equal to a small fixed portion of $\vol(\Delta^{n-1})$ (e.g.\ $1\%$). Then, to obtain the copula one has to estimate all the ratios $\frac{\vol(Q_{ij})}{\vol(\Delta^{n-1})}$ where 
$    Q_{ij} :=\{ x\in\Delta^{n-1}\ |\ s_i\leq f_{ret}(x, R)\leq s_{i+1} \text{ and } u_j\leq f_{vol}(x, \Sigma)\leq u_{j+1}\}$.

To compute these ratios, we leverage uniform sampling from $\Delta^{n-1}$~\cite{RbMel98}. 
Let us consider up- and down- (main) diagonal bands: we define the indicator as the ratio of the down-diagonal over the up-diagonal band. 
The indicator is the ratio of the mass of portfolios in the blue area over the mass of portfolios in the red one in Fig.~\ref{fig:cop2}. When the value of the indicator is smaller than $1$ then the copula corresponds to a normal period. Otherwise, it probably comes from a crisis period.

\subsection{Shock detection using real data}\label{sec:computations_on_data}

We now use two data sets from two different asset sections. First, we use the daily returns of 30 French industrial asset returns\footnote{\scriptsize\url{https://mba.tuck.dartmouth.edu/pages/faculty/ken.french/data_library.html}}. 
Second, we use the daily returns of 12 out of the top 100 cryptocurrencies, ranked by CoinMarketCap's\footnote{\scriptsize\url{https://coinmarketcap.com/}} market cap (cmc\_rank) on 22/11/2020, having the longest available history (Table~\ref{Tab:cryptos}). We compute the daily return for each coin using the daily close price obtained by CoinMarketCap, for several notable coins such as Bitcoin, Litecoin and Ethereum.

\begin{figure}[t]
    \centering
    \includegraphics[width=\linewidth]{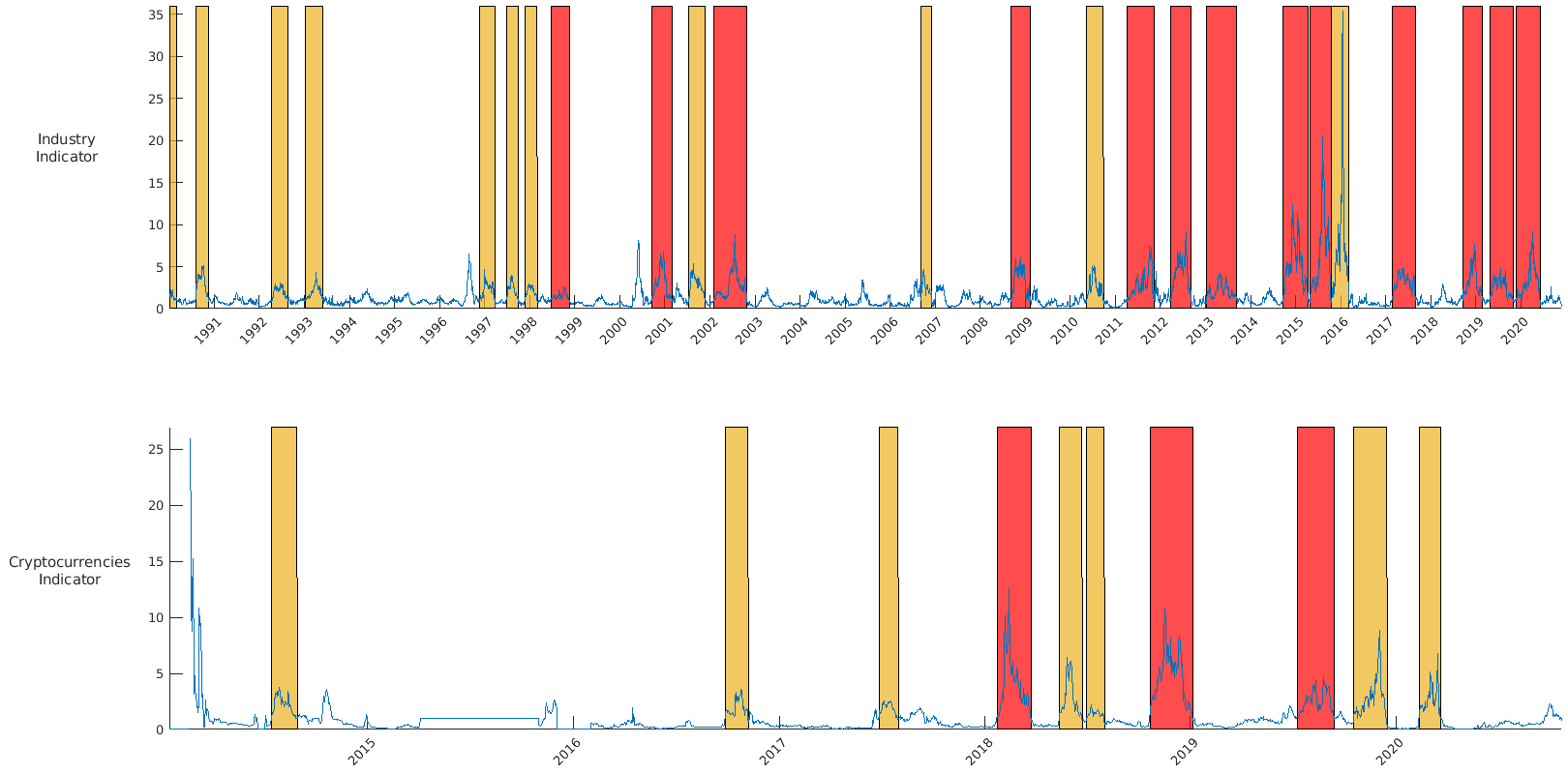}
    \caption{Warning (yellow) and Crises (red) periods detected by the indicator. Top for industry assets (1990-2020), bottom for cryptocurrencies (2014-2020).}
    \label{fig:industry_crypto_indicator}
\end{figure}

The indicator is estimated on copulae by drawing $500,000$ points. 
We compute the indicator per copula over a rolling window of $k=60$ days and with a band of $\pm 10\%$ with respect to the diagonal. When the indicator exceeds 1 for more than 60 days but less than 100 days, we report the time interval as a ``warning'' (yellow color); see Fig.~\ref{fig:WarningsCrises}. When the indicator exceeds~1 for more than 100 days we report the interval as a ``crisis'' (red); see Fig.~\ref{fig:industry_crypto_indicator}, \ref{fig:WarningsCrises}, and~\ref{fig:WarningsCrisesCryptos_vs_BTC}. 
The periods are more than 60 days long to avoid detection of isolated events whose persistence is only due to the auto-correlation implied by the rolling window. 

We compare results for industrial assets with the database for financial crises in Europe~\cite{ESRB17} from 1990 until 2020. 
The first warnings in 1990 correspond to the early 90's recession, 
the second crisis in 2000 to 2001 to the dot-com bubble burst, 
the warning and third crisis in 2001 and 2002 to the stock market downturn of 2002, 
and the fourth crisis in 2008 to 2009 corresponds to the sub-prime crisis.

Our cryptocurrencies indicator  
detects successfully the 2018 (great) cryptocurrency crash. The first shock event detected in 2018 (mid-January to late March) corresponds to the crash of nearly all cryptocurrencies, following Bitcoin's, whose price fell by about 65\% from 6 January to 6 February 2018, after an unprecedented boom in 2017. 
Intermediate warnings (mid-May to early August) should correspond to cryptocurrencies collapses (80\% from their peak in January) until September. 
The detected crash at the end of 2018 (November 2018 until early January 2019) corresponds to the fall of Bitcoin's market capitalization (below \$100 billion) and price by over 80\% from its peak, almost one-third of its previous week value. Finally, the detected event in early 2020 corresponds to the shock event due to covid-19.

\section{Exploring the dynamics of copulae}\label{sec:dynamic_copulas}

Several clustering methods confirm the indicator's reliability. Then, we model the structure of copulae using a novel regression model to detect shock events.

\begin{figure}[t]
    \centering
    \includegraphics[width=.5\linewidth]{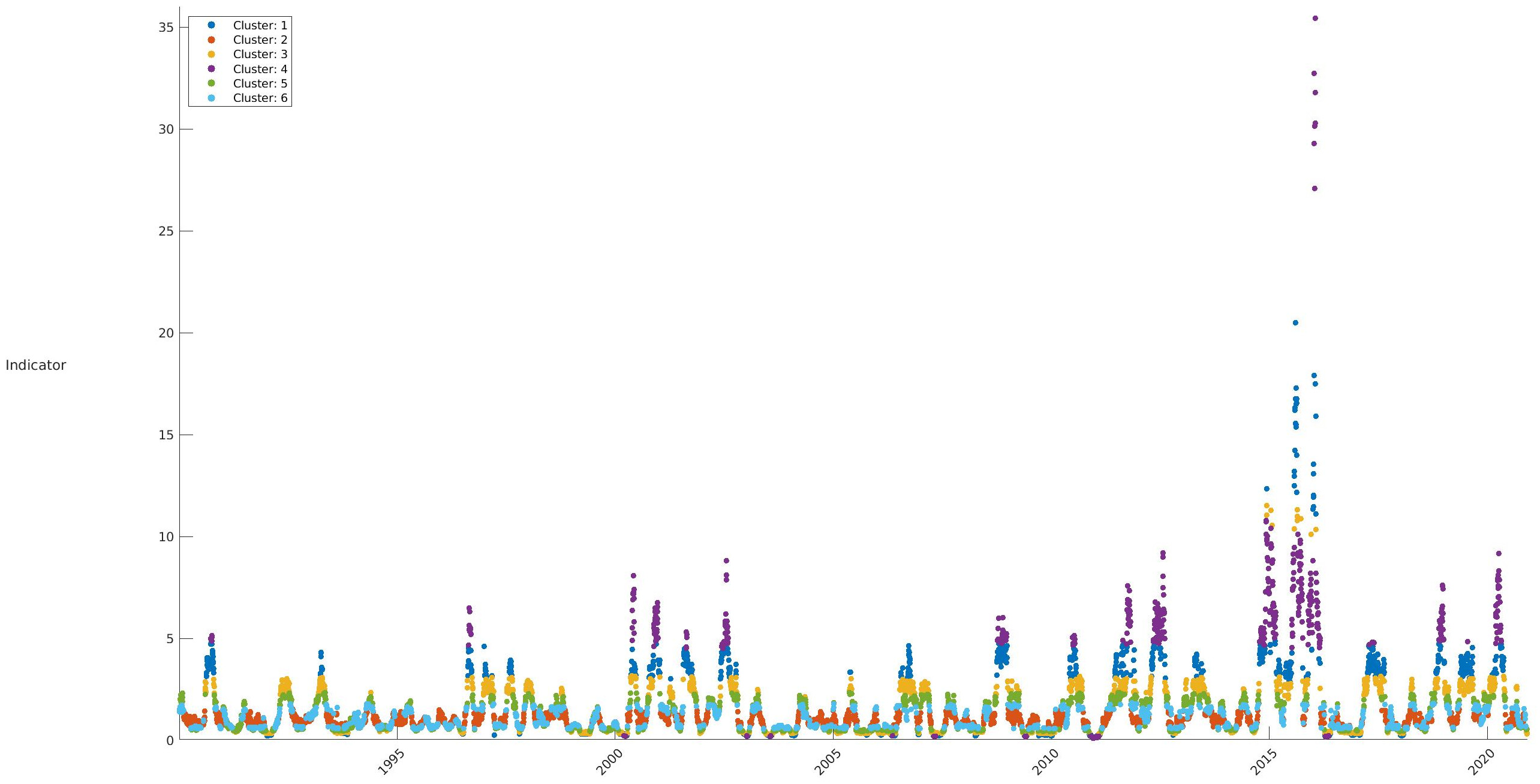}%
    \includegraphics[width=.5\linewidth]{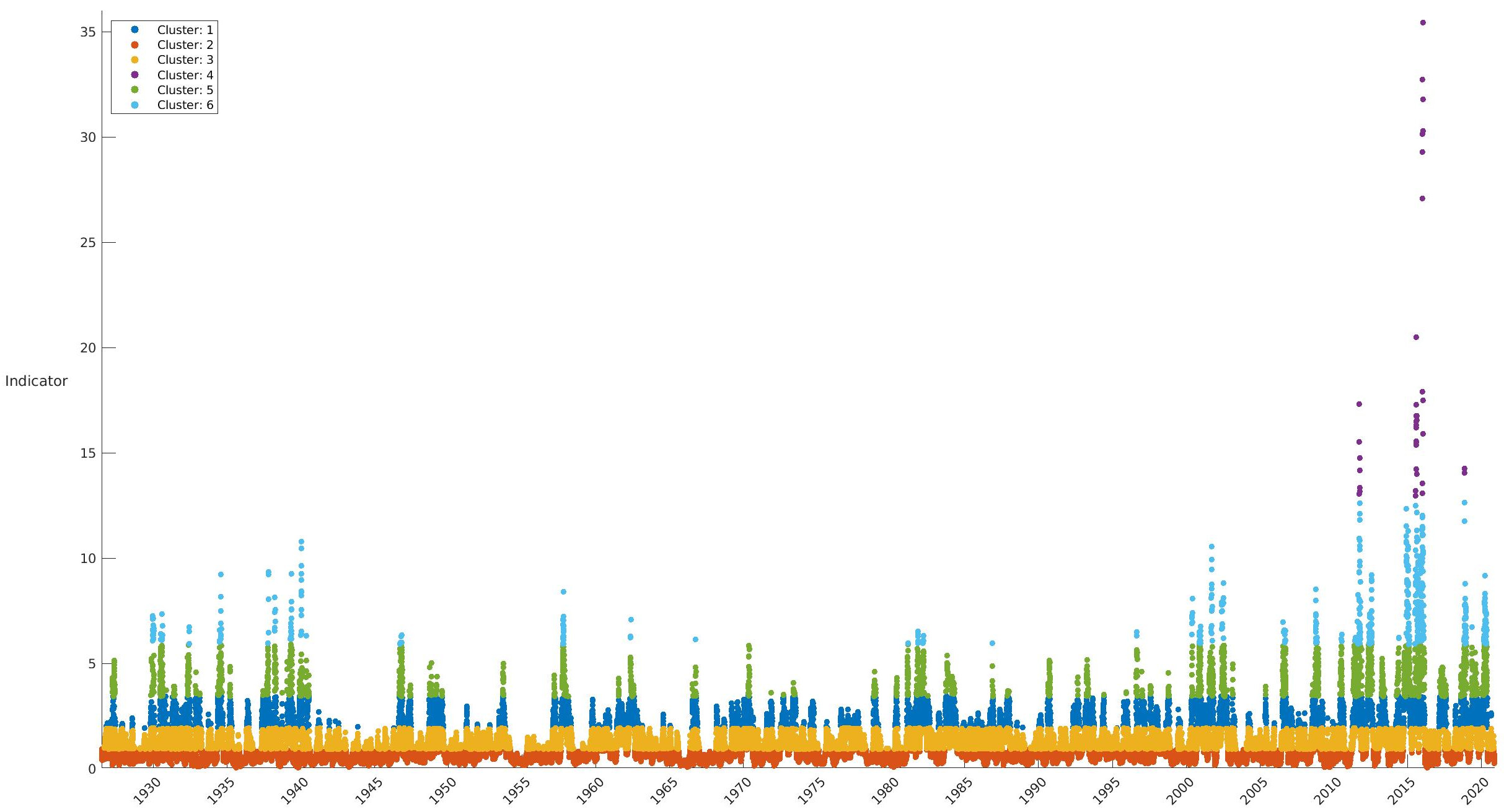}
    \caption{Left, spectral clustering ($k=6$) on EMD matrix. Right, k-medoids ($k=6$) on copulae features. Clusters appear to contain similar indicator values.}

    \label{fig:copulaeClusteringC6_clusters}
\end{figure}

\subsection{Clustering of copulae}

In order to further evaluate our results we clustered the copulae of the industry returns. Aiming to identify whether the copulae are able to distinguish different market states (normal, crisis and intermediate), as well as to validate the indicator, we experimented with clustering based on probability distributions distances. 

To cluster the probability distributions distances of the copulae, we computed a distance matrix ($D$) between all copulae using the earth mover's distance (EMD) \cite{yossi2000}. 
The EMD between two distributions is the minimum amount of work required to turn one distribution into the other. Here we use a fast and robust EMD algorithm, which appears to improve both accuracy and speed~\cite{pele2009}. Then, we apply spectral clustering \cite{Andrew2001}, a method to cluster points using the eigenvectors of the affinity matrix which we derive from the distance matrix, computed by the radial basis function kernel, replacing the Euclidean distance with EMD, where $A_{ij}=exp(-D_{ij}^2/2\sigma^2)$, and $\sigma$ is the standard deviation of distances.
Using the $k$ largest eigenvectors of the laplacian matrix, we construct a new matrix and apply k-medoids clustering by treating each row as a point, so as to obtain $k$ clusters. The results with $k=6$ and $k=8$ are shown on the indicators' values in Fig.~\ref{fig:copulaeClusteringC6_clusters}, \ref{fig:copulaeClusteringC6EMD}, and \ref{fig:copulaeClusteringC8EMD}. Clusters appear to contain copulae with similar indicator values. Crisis and normal periods are assigned to clusters with high and low indicator values respectively. Therefore, the clustering of the copulae is proportional to discretising the values of the indicator.

Other experiments included clustering on features generated form the copulas, based on the indicator. We generate vector representations for each copula using the rates between all the possible combinations of the indicators’ corners: for $U_L$, $U_R$ being the upper left and right corner of a copula respectively, and for $L_L$, $L_R$ the lower left and right corners, the vector representation is  $[\frac{U_L}{U_R}\frac{U_L}{L_L}\frac{U_L}{L_R}\frac{U_R}{L_L}\frac{U_R}{L_R}\frac{L_L}{L_R}]$. These representations allow us to use clustering, such as k-medoids. Results of the clustering also follow the values of the indicator as expected (Fig. ~\ref{fig:copulaeClusteringC6_clusters}, \ref{fig:copulaeClustering}).

\subsection{Modeling copulae}

We further explore the dynamics of copulae by modeling the mass distribution using a quadratic regression model. 

\begin{figure}[t]
 \begin{minipage}[h]{0.32\textwidth}
\includegraphics[width=\linewidth]{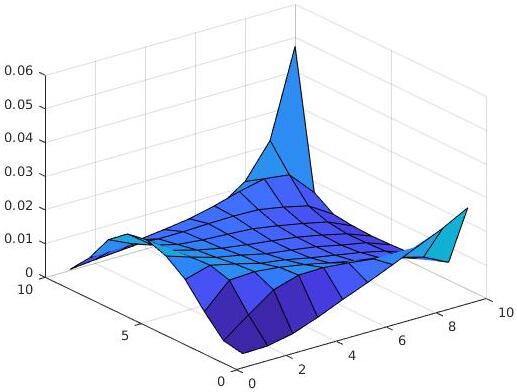}
\end{minipage}
 \begin{minipage}[h]{0.32\textwidth}
\includegraphics[width=\linewidth]{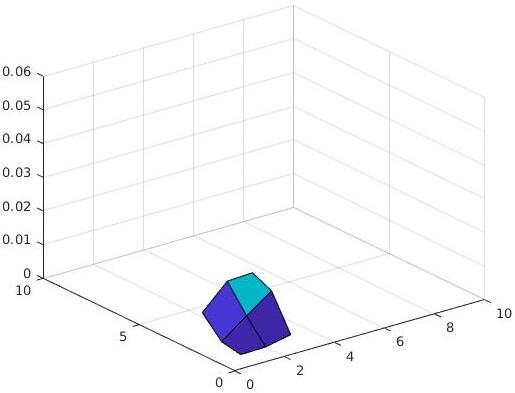}
\end{minipage}
\begin{minipage}[h]{0.32\textwidth}
\includegraphics[width=\linewidth]{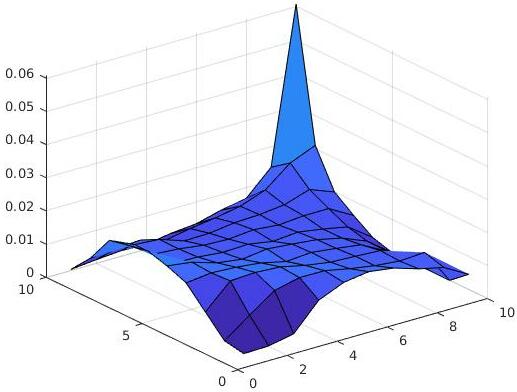}
\end{minipage}
\caption{Left: copula on 21/09/2014 using cryptocurrency returns. Middle: proportion of the mass of the left copula that our model uses as input. Right: copula that the model estimates. We trained our model with the industry asset returns. \label{fig:cop_estimation}}
\end{figure}

We compute the $10\times 10$ copulae of the industry data set. That is $N=49\, 689$ copulae in total while each consists of $10\times 10 = 100$ cells. Each cell has a value while they all sum up to $1$. For all copulae we pick a certain subset $S$ of $k$ cells, e.g.\ the $3\times 3$ left up corner as Fig.~\ref{fig:cop_estimation} illustrates; that is $|S|=9$. For each copula we represent the values of the $k$ cells that belong to $S$ as a vector $X\in\RR^k$. Thus, in total we get the vectors $X_1,\dots ,X_N$. Then, for each cell that does not belong to $S$ we fit a quadratic regression model. In particular, let $Y_{ij}\in\RR,\ i=1,\dots ,100-k,\ j=1,\dots ,N$ the value of the $i$-th cell in the $j$-th copula. Then, we define the following models,
\begin{equation}\label{eq:quadratic_models}
    \text{model } \mathcal{M}_i:\ \min\limits_{\Sigma_i\succeq 0} \sum_{j=1}^N(Y_j - X_j^T\Sigma_iX_j)^2,\quad i=1,\dots ,100-k ,
\end{equation}
where $\Sigma_i\succeq 0$ declares that the matrix $\Sigma_i$ is a positive semidefinite matrix. To solve the optimization problems in Equation~(\ref{eq:quadratic_models}) we use the {\tt matlab} implementation of the {\em Trust Region Reflective Algorithm}~\cite{Coleman99}, provided by function {\tt lsqnonlin()}.

To illustrate the efficiency and transferability of our model, we train it on industry asset returns and use it to detect the shock events in the cryptocurrency market from Section~\ref{sec:computations_on_data}; we use the copulae of the industry asset returns. For each copula the vector $X_i\in\RR^9$ corresponds to the $3\times 3$ left-down corner cell, as  Fig.~\ref{fig:cop_estimation} (middle) shows. We exploit the model to estimate the $10\times 10$ copula of each sliding window of the cryptocurrencies' returns. Finally, we compute the indicator of each estimated copula and plot the results in Fig.~\ref{fig:indicator_estimation}. Interestingly, the copulae that our model estimates suffice to detect all the past shock events that we also detect using the exact copulas, such as the 2018 cryptocurrency crash (see Sec.~\ref{sec:computations_on_data}), except the first warning period (mid-May) of 2018.
 

\subsubsection*{Acknowledgements} This research is carried out in the context of the project ``PeGASUS: Approximate geometric algorithms and clustering with applications in finance" (MIS 5047662) under call ``Support for researchers with emphasis on young researchers: cycle B" (EDBM103). The project is co-financed by Greece and the European Union (European Social Fund-ESF) by the Operational Programme Human Resources Development, Education and Lifelong Learning 2014-2020.
We thank Ludovic Calès for his precious guidance throughout this work. 


%
 \bibliography{references}
%


\appendix
\newpage
\section{Data}

\begin{table}[h!]
\centering
\begin{tabular}{l|c|c}
Coin & Symbol & Dates\\
\hline
Bitcoin	 & BTC & 28/04/2013 - 21/11/2020\\
Litecoin & LTC & 28/04/2013 - 21/11/2020\\
Ethereum & ETH & 07/08/2015 - 21/11/2020\\
XRP & XRP & 04/08/2013 - 21/11/2020\\
Monero & XMR & 21/05/2014 - 21/11/2020\\
Tether & USDT & 25/02/2015 - 21/11/2020\\
Dash & DASH & 14/02/2014 - 21/11/2020\\
Stellar & XLM & 05/08/2014 - 21/11/2020\\
Dogecoin & DOGE & 15/12/2013 - 21/11/2020\\
DigiByte & DGB  & 06/02/2014 - 21/11/2020\\
NEM & XEM & 01/04/2015 - 21/11/2020\\
Siacoin & SC & 26/08/2015 - 21/11/2020\\
\end{tabular}
\caption{Cryptocurrencies used to detect shock events in market.}
\label{Tab:cryptos}
\end{table}


\section{Crises indicator}

\begin{figure}[h!]
    \centering
    \includegraphics[width=\linewidth]{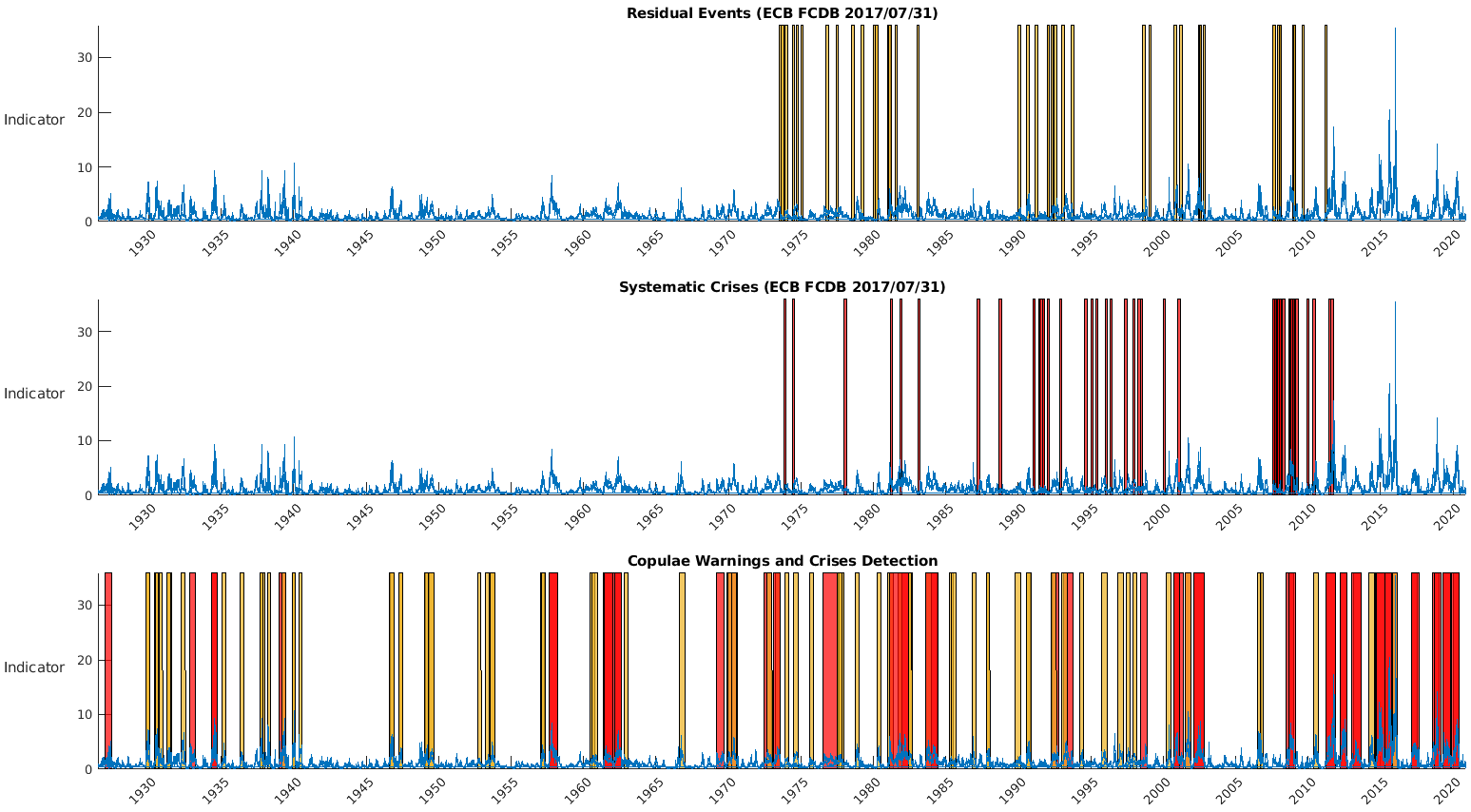}
    \caption{Warnings (yellow) and Crises (red) detected by indicator (bottom) for industry assets, against real residual events (top) and systematic crises (middle).}
    \label{fig:WarningsCrises}
\end{figure}

\begin{figure}[t]

\centering
\includegraphics[width=1.0\textwidth]{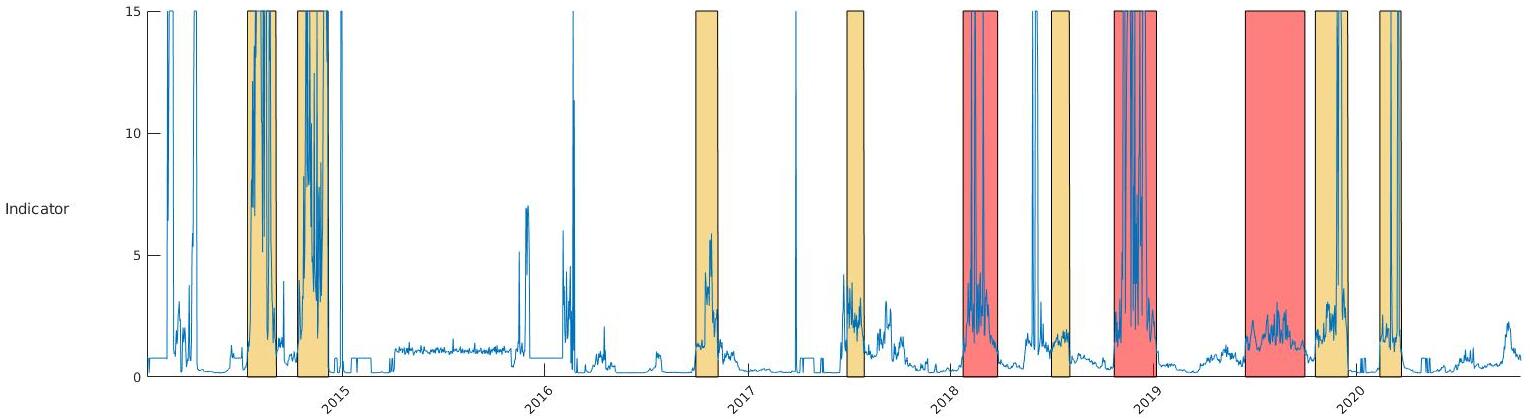}
\caption{The shock events we detect in the cryptocurrency market using the indicator from Equation~(\ref{eq:quadratic_models}). Note that we trained the model using the daily returns of the French industry assets. \label{fig:indicator_estimation}}
\end{figure}

\begin{figure}[t]
    \centering
    \includegraphics[width=\linewidth]{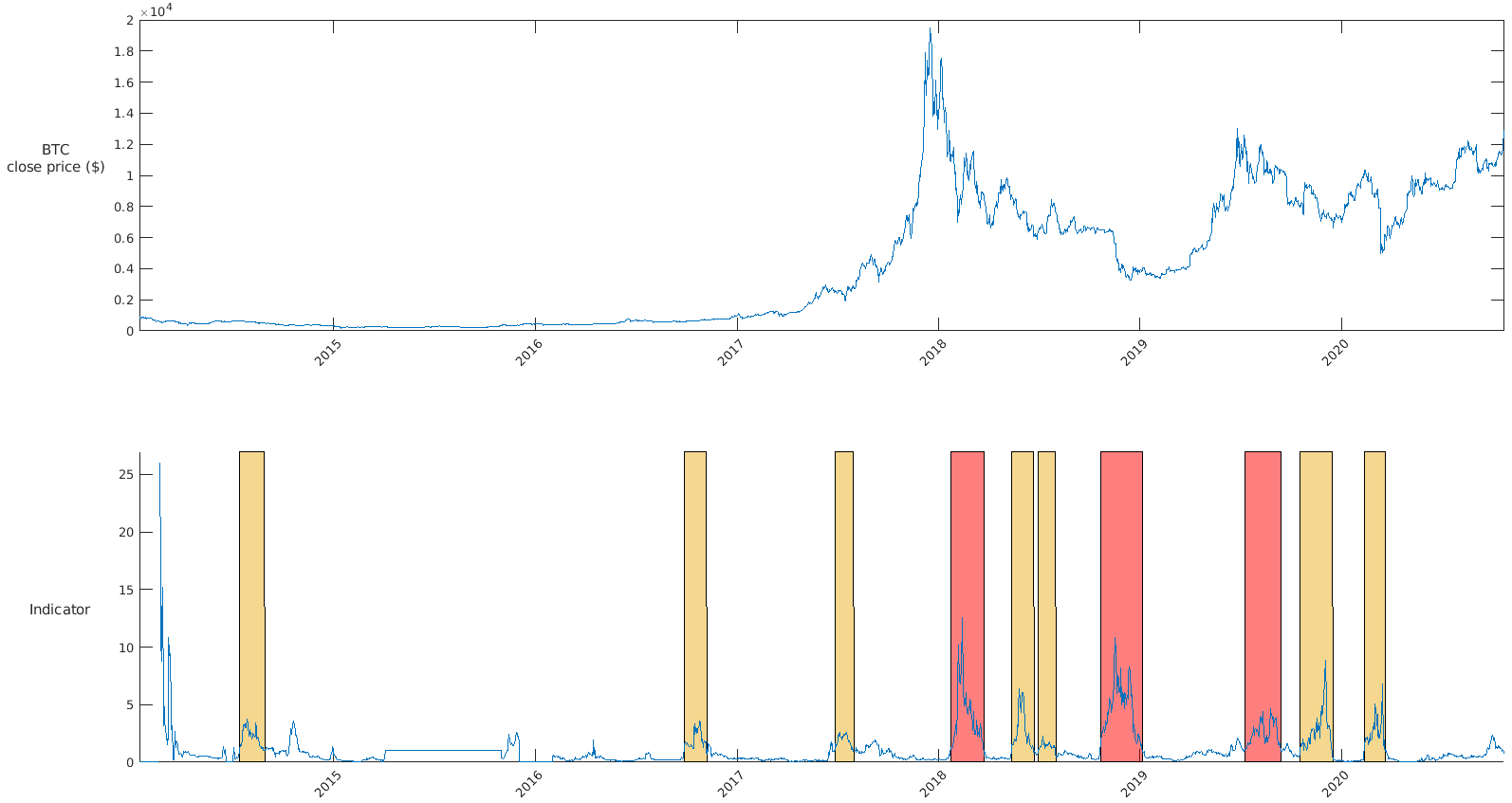}
    \caption{Warning (yellow) and Crises (red) periods detected by indicator (bottom) for cryptocurrencies against BTC daily close price (top).}
    \label{fig:WarningsCrisesCryptos_vs_BTC}
\end{figure}


\section{Clustering of Copulae}

\begin{figure}[h!]
    \centering
    \includegraphics[width=0.5\linewidth]{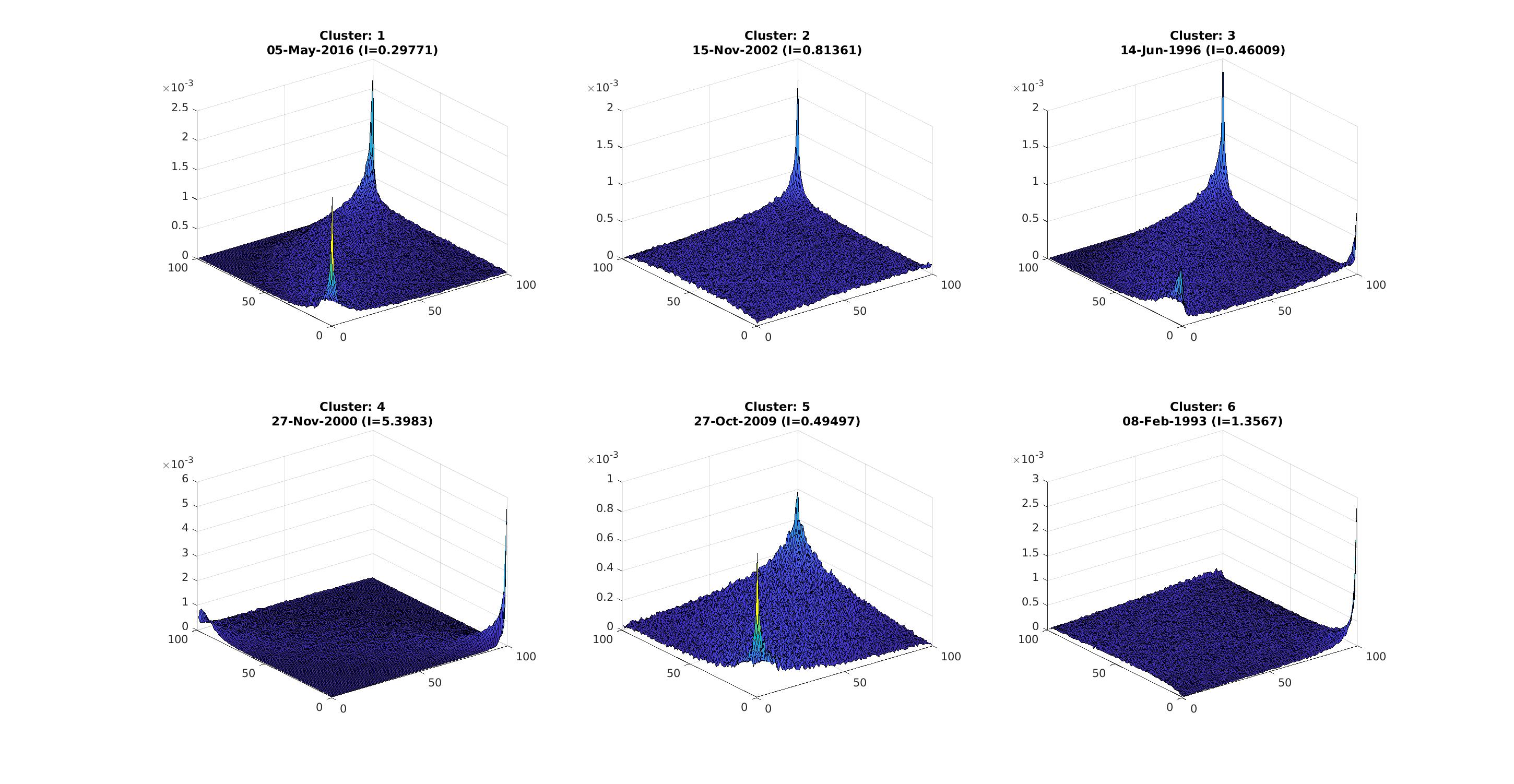}%
    \includegraphics[width=0.5\linewidth]{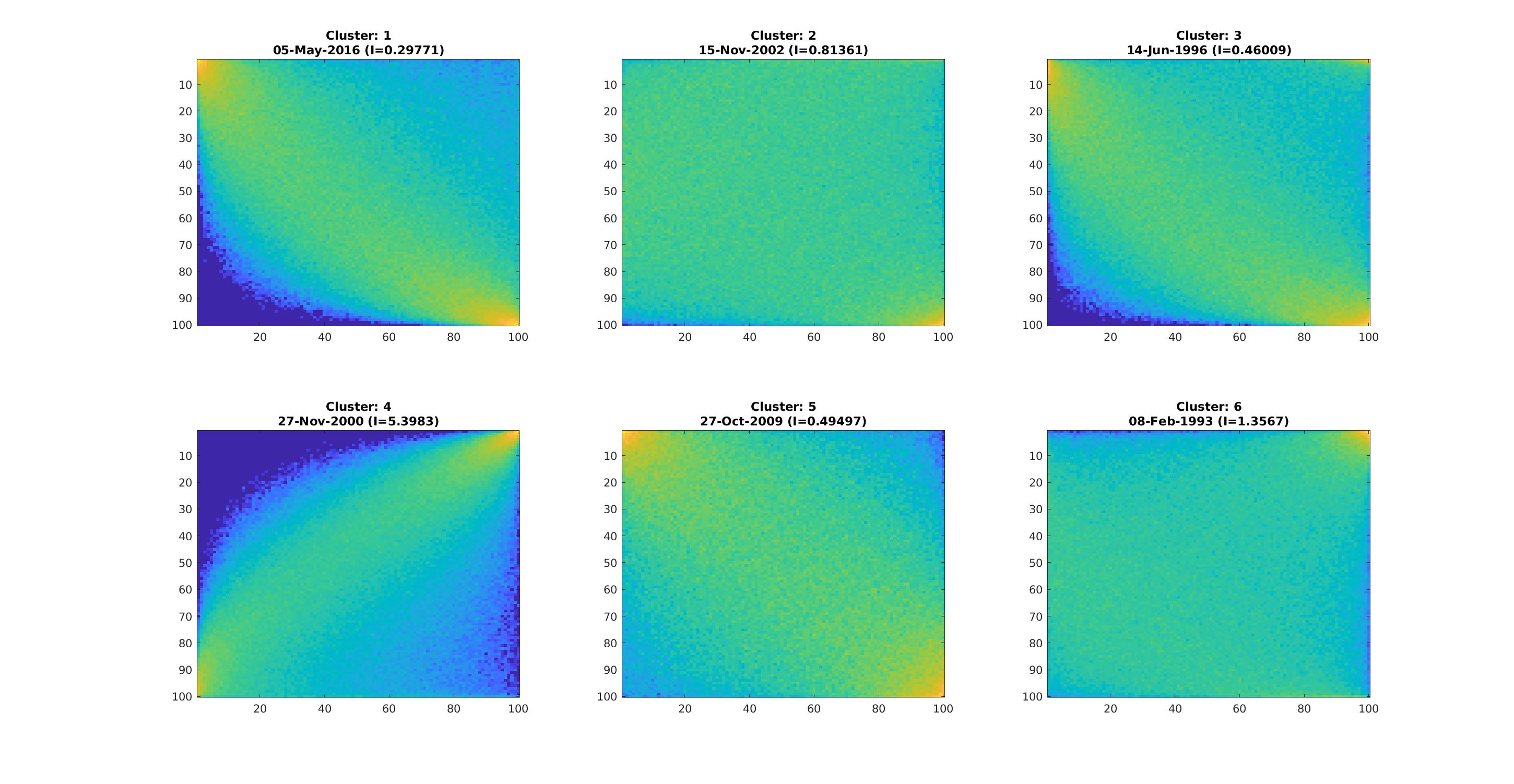}\\
    \includegraphics[width=.8\linewidth]{./clusters_C6_emd.jpg}
    \caption{Clustering of copulae using spectral clustering on EMD distances with $k=6$.}
    \label{fig:copulaeClusteringC6EMD}
\end{figure}

\begin{figure}[h!]
    \centering
    \includegraphics[width=0.5\linewidth]{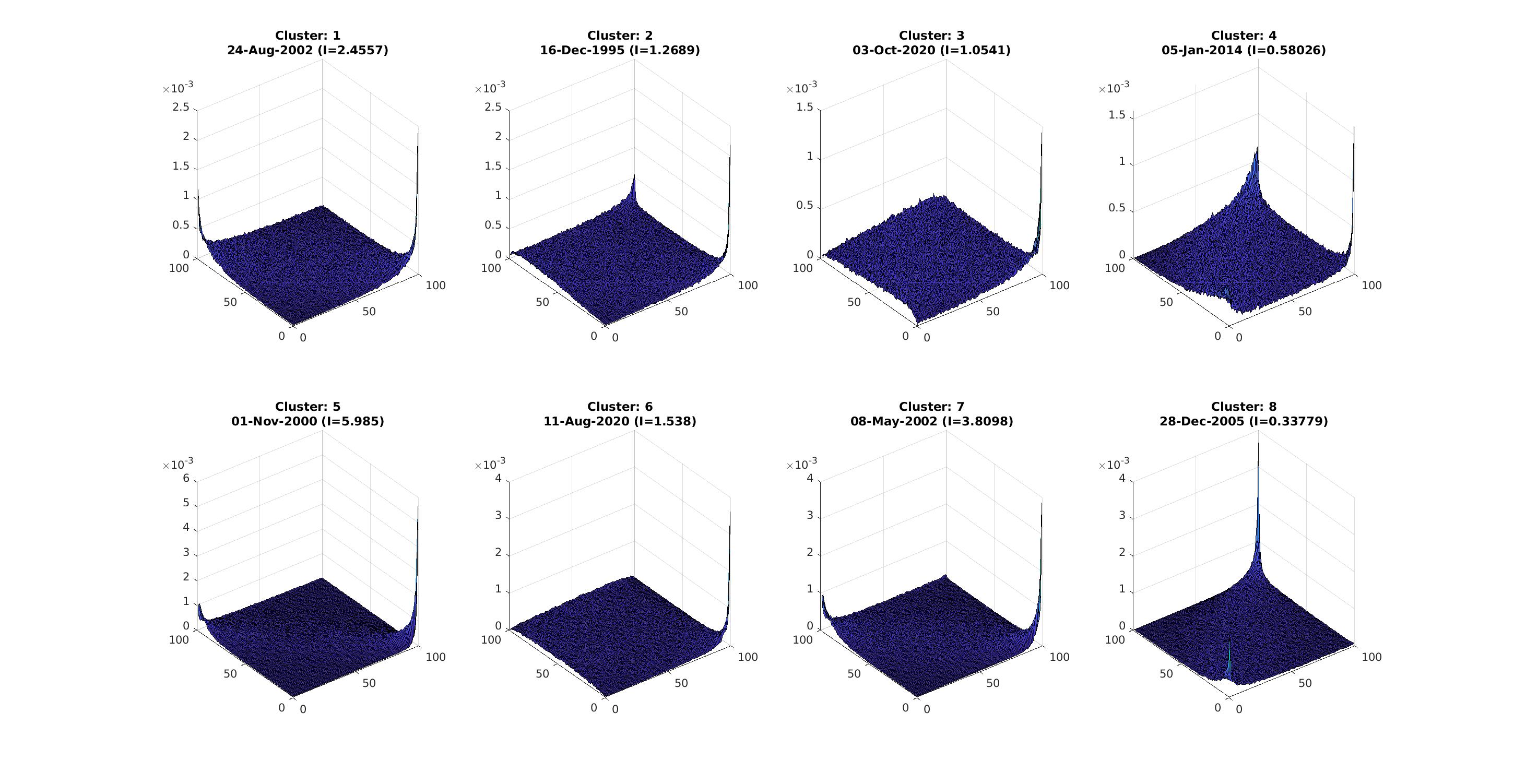}%
    \includegraphics[width=0.5\linewidth]{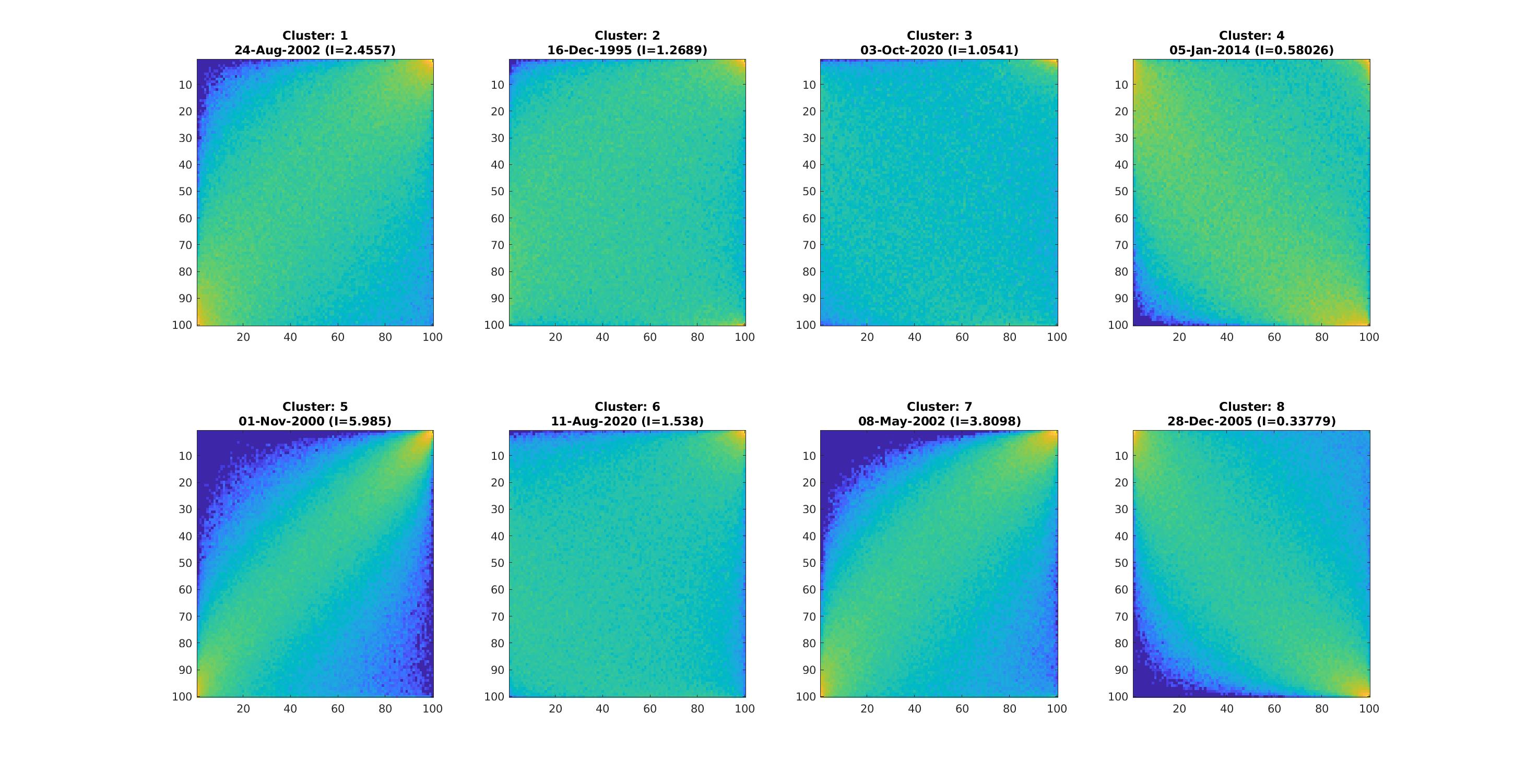}\\
    \includegraphics[width=.8\linewidth]{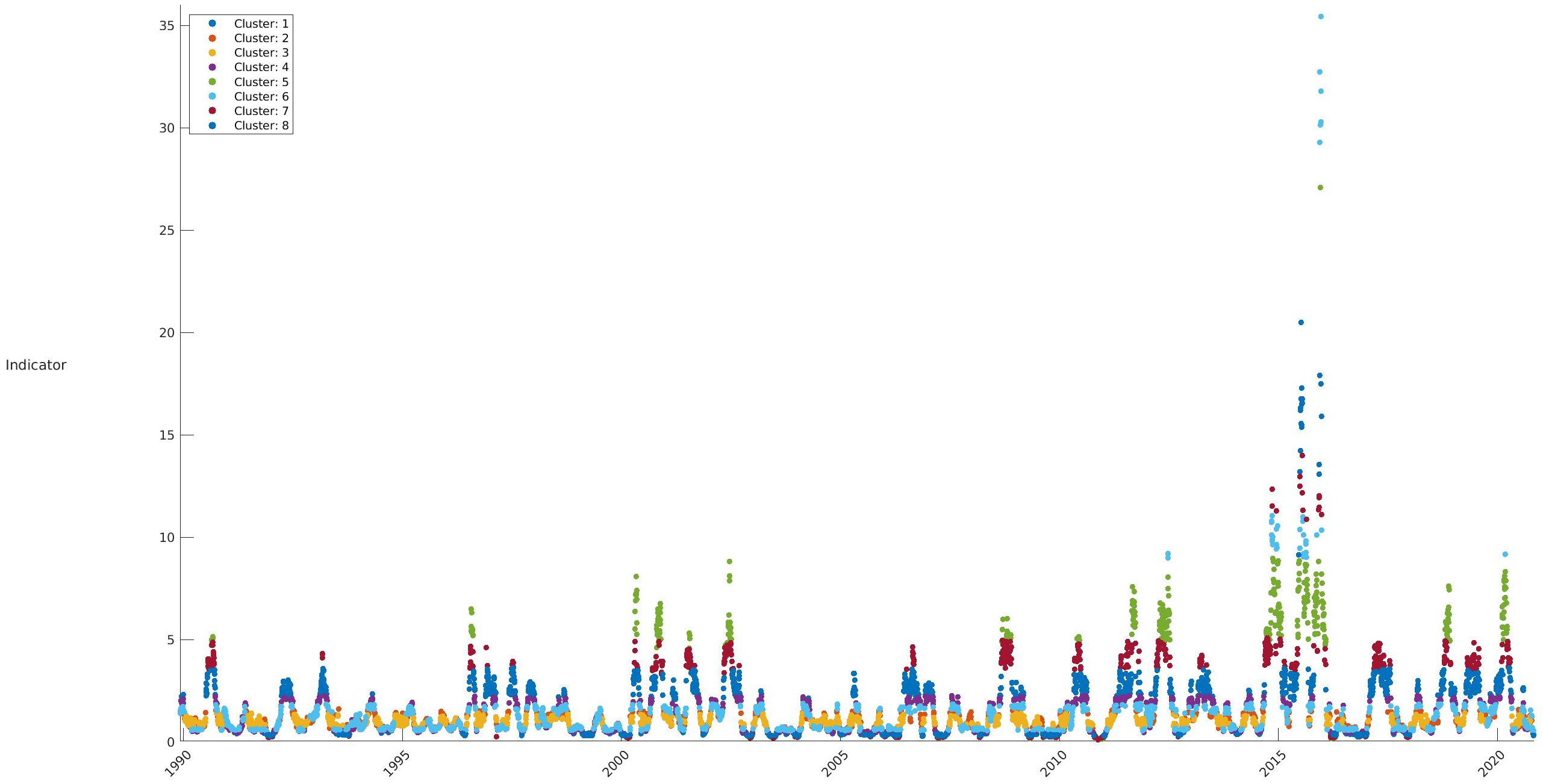}
    \caption{Clustering of copulae using spectral clustering on EMD distances with $k=8$.}
    \label{fig:copulaeClusteringC8EMD}
\end{figure}

\begin{figure}[h!]
    \centering
    \includegraphics[width=0.5\linewidth]{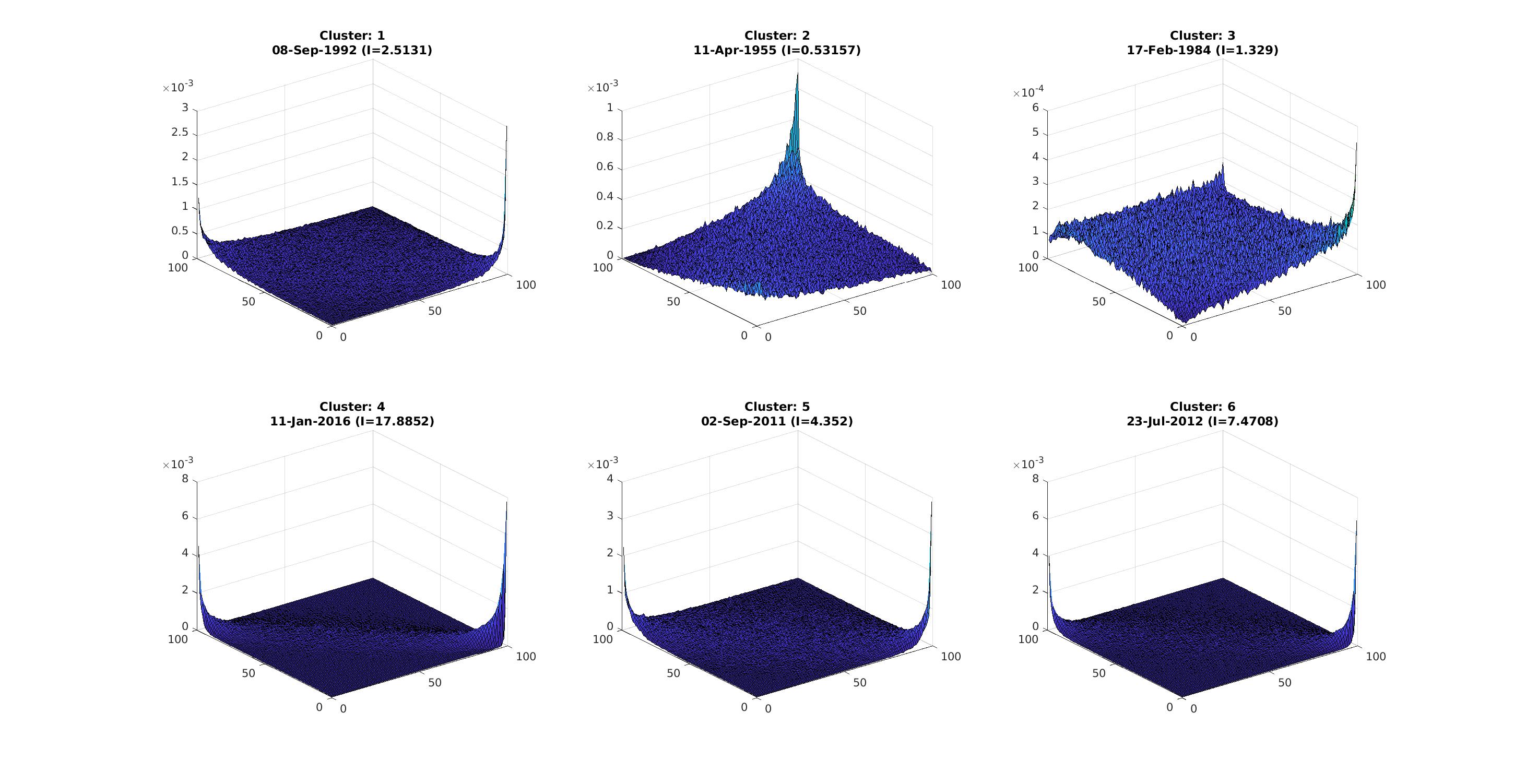}%
    \includegraphics[width=0.5\linewidth]{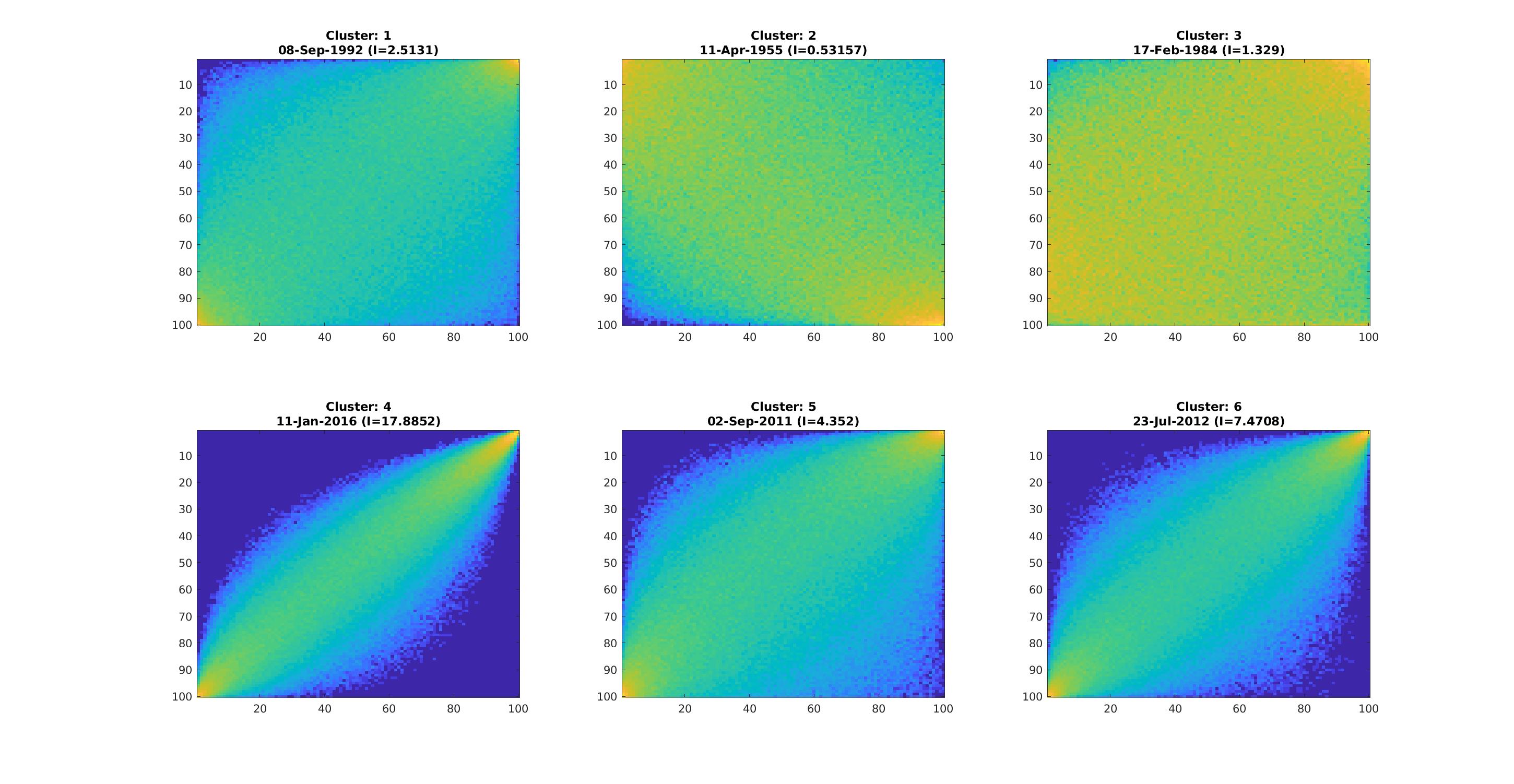}\\
    \includegraphics[width=.8\linewidth]{./clusters_C6_copulaFeats.jpg}
    \caption{Clustering using k-medoids on copulae features.}
    \label{fig:copulaeClustering}
\end{figure}


\end{document}